\documentclass{ws-ijmpcs}
\usepackage{bm}
\usepackage{graphicx}
\usepackage{color}
\newcommand{\ve}{\mathbf}
\usepackage{bbm}
\usepackage{amssymb}
\usepackage{amsmath}

\begin{document}

\title{TRANSPORT VIA CLASSICAL PERCOLATION AT QUANTUM HALL PLATEAU TRANSITIONS}

\author{MARTINA FL\"{O}SER}
\address{Institut N\'{e}el, CNRS and Universit\'{e} Joseph Fourier, B.P.
166, 25 Avenue des Martyrs, 38042 Grenoble Cedex 9, France}

\author{SERGE FLORENS}
\address{Institut N\'{e}el, CNRS and Universit\'{e} Joseph Fourier, B.P.
166, 25 Avenue des Martyrs, 38042 Grenoble Cedex 9, France}

\author{THIERRY CHAMPEL}
\address{Universit\'e Joseph Fourier Grenoble I / CNRS UMR 5493,
Laboratoire de Physique et Mod\'elisation des Milieux Condens\'es, B.P. 166, 
38042 Grenoble, France}

\maketitle

\begin{abstract}
We consider transport properties of disordered two-dimensional electron gases 
under high perpendicular magnetic field, focusing in particular on the peak longitudinal
conductivity $\sigma_{xx}^\mathrm{peak}$ at the quantum Hall plateau transition. 
We use a local conductivity model, valid at temperatures high enough such that quantum 
tunneling is suppressed, taking into account the random drift motion of the electrons in the 
disordered potential landscape and inelastic processes provided by electron-phonon 
scattering. A diagrammatic solution of this problem is proposed, which leads to 
a rich interplay of conduction mechanisms, where classical percolation effects play 
a prominent role.
The scaling function for $\sigma_{xx}^\mathrm{peak}$ is derived in the high temperature limit,
which can be used to extract universal critical exponents of classical percolation 
from experimental data. 
\end{abstract}

%\begin{history}
%\received{Day Month Year}
%\revised{Day Month Year}
%\end{history}

\ccode{PACS numbers: 73.43.Qt, 64.60.ah, 71.23.An}

\section{Introduction} 

The quantum Hall effect\cite{VonK1980,Prange} in two-dimensional electron gases (2DEG) 
follows from a disorder-induced localization process peculiar to the situation of 
large perpendicular magnetic fields $B$. While the formation of discrete Landau levels 
(LL) at energies $E_n=\hbar \omega_c \left(n+\frac{1}{2} \right)$ can account
for the existence of robust quantum numbers (with $\omega_c=|e|B/m^\star$ the cyclotron 
frequency, $e=-|e|$ the electron charge, $m^\star$ the effective mass, $\hbar$
Planck's constant divided by $2\pi$, and $n$ a positive integer), the existence of a 
macroscopic number of localized states in the bulk of 2DEG is an essential aspect of 
quantum Hall physics\cite{Hadju}.
Many questions are yet still open thirty years after the initial discovery: i)
for metrological purposes\cite{Cage2005}, which physical processes are limiting the plateau 
quantization of the Hall conductivity near the universal value $\sigma_{xy}=n e^2/h$?; 
ii) what is the nature of the localization/delocalization transition from one
plateau to the next\cite{Huckestein,Kettemann,EversRMP}, whereupon highly dissipative transport sets in?
Theoretically, the problem in its full complexity requires to understand the quantum
dynamics of electrons subject to the Lorentz force and random local electric
fields, possibly with the inclusion of dissipative processes such as
electron-electron and electron-phonon interaction, that are sensitive issues
when one considers transport properties.

So far, a lot of attention was turned towards the understanding of the delocalization
process in terms of a zero-temperature quantum percolation phase transition, which still 
remains a challenge for the theory\cite{Huckestein,Kettemann,EversRMP}, despite
intriguing experimental evidence from transport\cite{Wei1992,Li2010} and local
scanning tunneling spectroscopy\cite{HSWIM2008}. In this framework, the quantum
tunneling and interference of the guiding center trajectories within a complex percolation 
cluster allow dissipation to develop in a non-trivial way at the quantum Hall transition. 
Obviously, increasing temperature from absolute zero will generate inelastic processes 
limiting the coherence between saddle points of the disorder lanscape, so that
the quantum character of the transition becomes progressively irrelevant. In that case, a 
simpler quasiclassical transport theory becomes valid\cite{SH1994,PS1995,FS1995,FFC2011}, 
which incorporates the fast 
cyclotron motion with the slow guiding center drifting, and takes into account inelastic 
contributions to transport. The transport problem does not become however totally trivial, 
because classical percolation in the related advection-diffusion regime is still not fully
understood\cite{Isichenko}.

\begin{figure}[ht]
\includegraphics[scale=0.65]{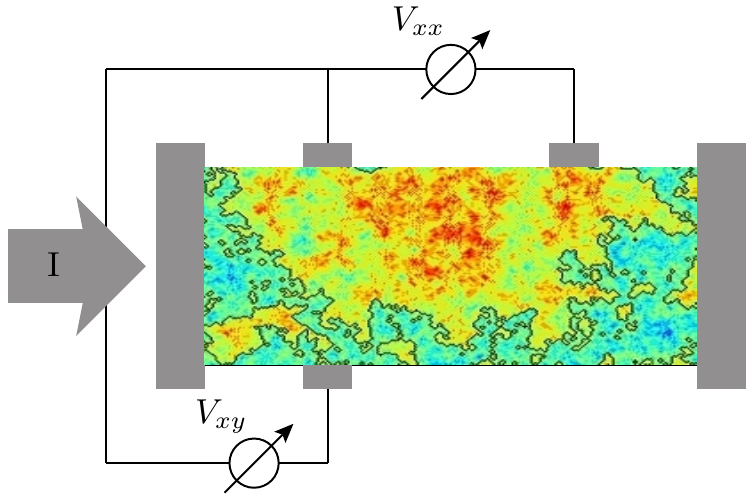}
 \includegraphics[width=0.60\linewidth]{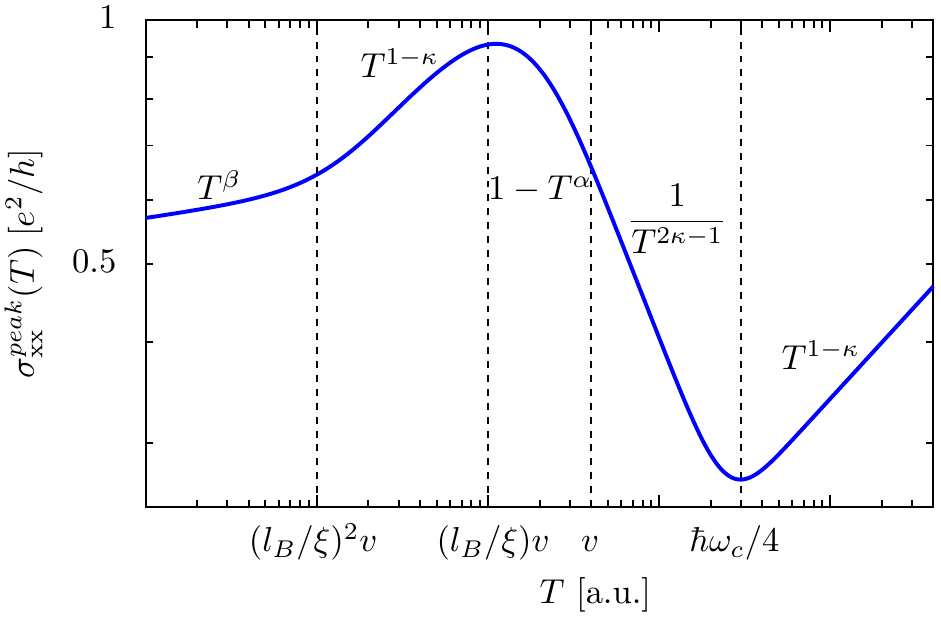}
\caption{Left: measurement of longitudinal $V_{xx}$ and Hall $V_{xy}$ voltages 
with applied current $I$ in a two-dimensional sample with percolating random charge
inhomogeneities. 
Right: sketch (on log-log scale) of the temperature dependence of the peak longitudinal
conductivity $\sigma_{xx}^\mathrm{peak}$ at the plateau transition. The existence of a 
hierarchy of energy scales (that are indicated by
dashed lines) results in several crossovers between universal power-laws, as decribed 
later in the text.}
\label{summary}
\end{figure}

The aim of the present paper is two-fold. 
First, we will show in Sec.~\ref{sec:crossovers} that high mobility samples
display a very rich temperature behavior for the peak longitudinal conductivity 
$\sigma_{xx}^\mathrm{peak}$ (at the plateau transition), leading to a complex succession of 
transport crossovers with universal powerlaws, see Fig.~\ref{summary}.
Second, we will present in Sec.~\ref{sec:diag} a general
diagrammatic formalism\cite{FFC2011} allowing to compute dissipative transport dominated
by classical percolation effects in quantum Hall samples such as depicted in 
Fig.~\ref{summary}.
In particular, we will be able to give strong support to a previously 
conjectured\cite{SH1994,PS1995,FS1995,Isichenko} critical exponent $\kappa=10/13$ 
for the peak longitudinal conductivity $\sigma_{xx}^\mathrm{peak}$ in the high
temperature regime of the plateau transition. An universal scaling function describing 
the crossover for temperatures near the cyclotron energy (see the V-shaped part of the 
curve in Fig.~\ref{summary}) will be computed, giving a way to extract $\kappa$ 
from experiments.

\section{Classically percolating transport at the plateau transition}
\label{sec:crossovers}

\subsection{Local conductivity model}

The starting point of transport calculations in the high temperature
regime of the quantum Hall effect is a purely classical
model\cite{SH1994,FFC2011}, where the continuity equation 
$\bm\nabla\cdot\ve{j}=0$ ({\it i.e.} the continuum version of Kirchoff's law) 
is solved from the microscopic knowledge of Ohm's law
$\ve{j}(\ve{r})=\hat\sigma(\ve{r})\ve{E}(\ve{r})$, defining here the local 
conductivity tensor\cite{GV1994,CFC2008} that relates the local electric field 
to the local current density. Due to the existence of several energy scales, 
the disorder-induced spatial variations of $\hat\sigma(\ve{r})$ have a strong
temperature dependence, which in turn affects the macroscopic transport properties. 
Drastic simplifications occur in the regime of high magnetic
fields\cite{CFC2008,CF2009}, where the combination of Lorentz force and local electrostatic 
potentials induces a slow drift motion in the direction orthogonal to the
crossed magnetic and local electric fields. This vindicates the first simplification 
of the conductivity tensor
\begin{equation}
\label{condTensor}
\hat\sigma(\ve{r})=
\begin{pmatrix}
\sigma_0 & -\sigma_H(\ve{r})\\
\sigma_H(\ve{r}) & \sigma_0
\end{pmatrix},
\end{equation}
where $\sigma_0$ encodes dissipative processes such as electron-phonon
scattering, which will be assumed to be uniform in the bulk of the sample,
and $\sigma_H(\ve{r})$ is the local Hall component, whose spatial dependence
originates from charge density fluctuations due to disorder in the sample.
We will discuss below the various regimes that can be expected for $\sigma_H(\ve{r})$
depending on the range of temperature $T$.
For this purpose, we need to introduce the energy scales associated with the local disorder 
potential $V({\bf r})$, and we define its typical amplitude $v=\sqrt{\big<[V({\bf r})]^2\big>}$
and correlation length $\xi$. We will assume throughout that the
disorder is smooth at the scale of the magnetic length $l_B=\sqrt{\hbar/eB}$ 
($l_B=8$nm at $B=10$T), so that $l_B/\xi$ is a small parameter. 
%For instance, first quantum corrections due to wavefunction spread occur at the energy scale
%$(l_B/\xi)v$, while quantum tunneling sets in at the smaller scale $(l_B/\xi)^2
%v$, see Refs.~\refcite{CFC2008,CF2009}. 
In what follows, a centered Gaussian distribution will be considered for the local 
electrostatic potential $V(\ve{r})$. 
The local conductivity model introduced here is valid at temperatures high
enough so that phase-breaking processes, such as electron-phonon 
scattering, occur on length scales that are shorter than the typical variations of 
disorder. However, quantum mechanics may still be important to determine
the microscopics of the conductivity tensor, as we argue below.

\subsection{Percolation effects in quantum Hall transport: phenomenology}
\label{phenomenology}

The occurence of percolation effects in the quantum Hall regime can be
understood already from a quasiclassical perspective. In the high magnetic field
limit, cyclotron and guiding center motions fully decouple, giving rise to
Landau quantization on one hand, and to mainly closed trajectories of the
guiding center on the other hand, that follow the equipotentials of the disorder
landscape. Intuitively, the electrical current contributing to macroscopic
transport will thus follow a percolation backbone.
The crucial role of inelastic processes, controlled by the longitudinal component 
$\sigma_0$ in Eq.~(\ref{condTensor}), can be understood by the fact that such 
current-carrying extended states must pass through many saddle-points on the disorder
landscape. However, the drift velocity associated to the guiding center identically 
vanishes at these points, so that having a finite $\sigma_0$ is essential to connect 
the different valleys of the potential profile. The technical difficulty lies in 
evaluating the macroscopic conductivity in the limit where $\sigma_0$ is much smaller than the
amplitude variations of the Hall component $\sigma_H(\ve{r})$, but yet does not
fully vanish. This regime cannot simply be accessed from the $\sigma_0\to0$
limit, because the transport equation becomes singular. The strategy developped in
Ref.~\refcite{FFC2011} and Sec.~\ref{sec:diag} will be to extrapolate from high orders 
of the perturbatively controlled $\sigma_0\to\infty$ expansion to the case of small 
dissipation.

Assuming that a critical state is established in the small $\sigma_0$ limit due
to the scale invariant nature of the percolation backbone, one can infer from 
dimensional analysis that the macroscopic longitudinal conductivity scales 
as\cite{SH1994,FFC2011}:
\begin{equation}
\sigma_{xx}\propto \sigma_0^{1-\kappa}
[\big<\sigma_H^2\big>-\big<\sigma_H\big>^2]^{\kappa/2},
\label{eq:scaling}
\end{equation}
where $\kappa$ is a non-trivial exponent previously
conjectured\cite{SH1994,PS1995,FS1995,Isichenko} to be $\kappa=10/13$, see
Sec.~\ref{sec:diag} for a diagrammatic approach to this result.
Based on simple microscopic arguments for the local Hall conductivity $\sigma_H(\ve{r})$
that we introduce now, it is possible to understand from Eq.~(\ref{eq:scaling})
various transport regimes that are relevant for quantum Hall systems.
In all what follows, we will assume that electron-phonon processes dominate in
the longitudinal component\cite{Feng}, leading to the temperature dependence 
$\sigma_0(T)\propto T$.

\subsection{A hierarchy of transport crossovers}
\label{hierachy}

\subsubsection{Fully classical regime: $\hbar\omega_c\ll T$}
At temperatures higher than the cyclotron energy, both cyclotron and drift
motions are classical, so that the classical Hall's law prevails: 
$\sigma_H(\ve{r})= (e/B) n(\ve{r})$, with $n(\ve{r})$ the local electronic
density, which undergoes smooth spatial fluctuations in case of high mobility
samples\cite{SH1994}. In relatively clean samples, the amplitude 
$v$ of disorder fluctuations remain small compared to the classical cyclotron 
energy, so that the Hall conductivity follows at first order the spatial variations 
of the local potential: 
\begin{equation}
\sigma_H(\ve{r})= \frac{e n}{B}+ A V(\ve{r}).
\label{sigmaH_classical}
\end{equation}
with $n$ the total electron density and $A$ a constant to be determined below.
Thus, for a Gaussian distributed disorder, the local Hall conductivity displays Gaussian 
fluctuations and is weakly dependent of temperature. Using the percolation
Ansatz~(\ref{eq:scaling}) for the macroscopic longitudinal conductivity, we
find:
\begin{equation}
\sigma_{xx}\propto v^\kappa T^{1-\kappa}\propto T^{3/13},
\label{sigmaxx_classical}
\end{equation}
which shows already a first non-trivial behavior in
temperature\cite{PS1995,FFC2011} connected to classical percolation, where the
conductivity mildly decreases as temperature is lowered, see also
Fig.~\ref{summary}.

\subsubsection{Formation of Landau levels: $v\ll T\ll \hbar\omega_c$}

As temperature crosses the cyclotron energy, Landau levels start to emerge,
and the local density is given by Pauli's principle: 
$\sigma_H(\ve{r})= {\displaystyle \frac{e^2}{h}} {\displaystyle\sum_{m=0}^{\infty}} n_F[E_m-V(\ve{r})-\mu]$
with $\mu$ the chemical potential and $n_F(E)=1/(e^{E/T}+1)$ the
Fermi-Dirac distribution (we set Boltzmann's constant $k_B=1$ in what follows). 
We will neglect spin effects for simplicity in what follows (Landau levels
are assumed spin non-degenerate).
In the considered temperature range $v\ll T$, the Fermi
distribution can be linearized, which leads to Eq.~(\ref{sigmaH_classical}) 
with $A=(e^2/h)(\hbar \omega_c)^{-1}$ in the case $\hbar\omega_c\ll T$ considered
previously, and more generally to:
\begin{equation}
\sigma_H(\ve{r})= \frac{e n}{B} + \frac{e^2}{h}\sum_{m=0}^\infty n_F'(E_m-\mu) V(\ve{r}).
\label{sigmaH_LL}
\end{equation}
The local conductivity remains Gaussian, but acquires now an extra temperature
dependence from the Fermi function, which can be illustrated in the case
of the plateau $\nu\to\nu+1$ transition, which leads for $T\ll \hbar\omega_c$ to
\begin{equation}
\sigma_H^\mathrm{peak}(\ve{r})= \frac{e n}{B} + \frac{e^2}{h}\frac{1}{4T} V(\ve{r}).
\end{equation}
Using the percolation Ansatz~(\ref{eq:scaling}), we find in the considered
temperature range:
\begin{equation}
\sigma_{xx}^\mathrm{peak}\propto \left(\frac{v}{T}\right)^\kappa T^{1-\kappa}\propto
\frac{1}{T^{2\kappa-1}}\propto \frac{1}{T^{7/13}},
\label{sigmaxx_LL}
\end{equation}
so that the peak longitudinal conductivity strongly increases below
$T\lesssim\omega_c/4$ (this crossover scale, as well as the complete scaling
function will be determined in Sec.~\ref{sec:diag}), see also
Fig.~\ref{summary}.

\subsubsection{Two-fluids regime: $(l_B/\xi)v\ll T\ll v$}
The peak longitudinal conductivity cannot diverge at vanishing temperature,
and the law~(\ref{sigmaxx_LL}) must be cut-off by additional physical processes.
Indeed, by further lowering the temperature, the Fermi distributions becomes 
sharp at the scale $T\ll v$, and the local Hall conductivity $\sigma_H(\ve{r})$
now assumes rapid spatial variations between quantized values $\nu e^2/h$ and $(\nu+1)e^2/h$, 
with $\nu$ the filling factor. The local conductivity model now reads
\begin{equation}
\sigma_H(\ve{r})= \frac{e^2}{h}\nu + \frac{e^2}{h}\Theta[V(\ve{r})+\mu-E_\nu],
\label{sigmaH_twofluids}
\end{equation}
introducing the step function $\Theta$. The transport properties of this two-fluids 
model were considered extensively in previous works\cite{DR1994}. It was found using duality 
arguments that
the local conductivity Eq.~(\ref{sigmaH_twofluids}) leads in the (unphysical)
limit of zero temperature to an exact value for the longitudinal peak
conductivity in the $\sigma_0\to0$ limit: $\sigma_{xx}^\mathrm{peak}=e^2/h$. This result
would seem at first sight at odds with the scaling Ansatz~(\ref{eq:scaling}),
which predicts a powerlaw vanishing of $\sigma_{xx}$ at small $\sigma_0$.
On mathematical grounds, the model Eq.~(\ref{sigmaH_twofluids}) is quite peculiar 
in the sense that the fluctuations of the Hall conductivity 
$[\big<\sigma_H^2\big>-\big<\sigma_H\big>^2]$ are actually diverging at the peak
value, invalidating the Ansatz. Yet, the existence of a finite and universal
value $\sigma_{xx}^\mathrm{peak}=e^2/h$ seems still physically surprising from
the argumentation given in Sec.~\ref{phenomenology}, where we argued that
the fully opened current lines at $\sigma_0$ have a vanishing drift velocity
at the saddle points of disorder. However, for such bimodal distribution of the
local Hall conductivity Eq.~(\ref{sigmaH_twofluids}) and in contrast to any
continuous conductivity distribution, the drift velocity does not vanish anymore 
at the saddle points, which allows to establish a macroscopic current even in
the absence of dissipation mechanisms. This simple argument allows to understand
why the percolation scaling Ansatz~(\ref{eq:scaling}) does not apply to the 
two-fluids model of Dykhne and Ruzin\cite{DR1994}. However, we will see below that other processes
invalidate the model Eq.~(\ref{sigmaH_twofluids}) in the limit of zero
temperature. Mreover, we can also infer how the ``exact'' value $e^2/h$ is
approached from above. Indeed, the sharp Fermi function in
Eq.~(\ref{sigmaH_twofluids}) is always smeared on the scale $T$, recovering a
continuous (but strongly non-Gaussian) distribution, leading likely to power-law 
deviations from the exact zero-temperature result $e^2/h$:
\begin{equation}
\sigma_{xx}^\mathrm{peak}= \frac{e^2}{h} - B T^\alpha 
\label{sigmaxx_twofluids}
\end{equation}
with a new critical exponent $\alpha>0$ that is to our knowledge still unknown,
and $B$ some constant.
The fact that the peak longitudinal conductivity levels off at low temperatures
towards values close (but not strictly equal) to $e^2/h$ has been noted from
experimental data\cite{DR1994}, see also Fig.~\ref{summary}.

\subsubsection{Wavefunction corrections: $(l_B/\xi)^2 v\ll T\ll (l_B/\xi) v$}
The low-temperature two-fluids conductivity model Eq.~(\ref{sigmaH_twofluids})
relies on the high magnetic field limit, and is strictly speaking only correct
in the limit $l_B\to0$. However, quantum corrections will occur for finite $l_B/\xi$ 
due to the fact that the electronic wavefunctions are not
infinitely sharp transverse to the guiding center motion, but rather spread on
the scale of the magnetic length $l_B$. For this reason, the local Hall
conductivity $\sigma_{xx}(\ve{r})$ will not undergo infinitely sharp steps
from a quantized value to the next as in Eq.~(\ref{sigmaH_twofluids}), but rather rapid 
but smooth rises on the scale $l_B$, see Refs.\refcite{CFC2008,CF2009}. 
Because the wavefunctions extend
transversely in a Gaussian manner, the resulting form of the Hall conductivity is 
easily understood (here for the lowest Landau level):
\begin{equation}
\sigma_H(\ve{r})= \frac{e^2}{h}+ \frac{e^2}{h} \int \frac{d^2 \ve{R}}{\pi l_B^2} 
\; \Theta\left[V(\ve{R})-\mu-E_\nu\right] e^{-(\ve{r}-\ve{R})^2/l_B^2}.
\label{sigmaH_waveft}
\end{equation}
Clearly, the two-fluid model Eq.~(\ref{sigmaH_twofluids}) is recovered in the
limit $l_B/\xi\to0$, but for a more realistic smooth disorder, the correlation length
$\xi$ does not exceed a few hundreds of nanometers. In that case, the sharp
step in Eq.~(\ref{sigmaH_twofluids}) is smoothened whenever the new energy scale
$(l_B/\xi) v$ sets in. Interestingly, we recover now a continuous conductivity
distribution where the percolation Ansatz~(\ref{eq:scaling}) should apply. 
Because the spatial fluctuations of the local Hall conductivity are no more
controlled by temperature, we can infer without detailed calculation the
following powerlaw for the peak longitudinal conductivity:
\begin{equation}
\sigma_{xx}^\mathrm{peak} \propto T^{1-\kappa}\propto T^{3/13}.
\label{sigmaxx_waveft}
\end{equation}
Thus the peak conductivity should {\it decrease} again by cooling the sample to
very low temperatures, as evidenced experimentally\cite{Wei1992}, see also 
Fig.~\ref{summary}.

\subsubsection{Onset of quantum tunneling: $T\ll (l_B/\xi)^2 v$}
By further cooling towards the limit of zero temperature, a new energy scale 
$(l_B/\xi)^2 v$ emerges, associated to quantum tunneling at the saddle
points\cite{CF2009}.
For high mobility samples, one can assume that transport remains incoherent 
between the widely separated saddle points, so that quantum interference
effects can be neglected, and the local conductivity model Eq.~(\ref{condTensor})
still applies (if not, non-local effects in the spirit of
Ref.~\refcite{Baranger} must be accounted for). Here, the precise form of
the local conductivity tensor is not yet fully understood, although a quasilocal
approach that incorporates quantum tunneling can be developped\cite{CF2009}.
For this reason, the precise scaling form of the peak longitudinal conductivity
is still unknown in this regime, although a slower decrease than
Eq.~(\ref{sigmaxx_waveft}), leading to a kink at $T=(l_B/\xi)^2 v$, can be
expected, due to the onset of the quantum processes allowing to transfer
electrons above the saddle points: 
\begin{equation}
\sigma_{xx}^\mathrm{peak} \propto T^\beta \;\;\;\;\; 0<\beta<3/13.
\label{sigmaxx_tunnel}
\end{equation}
Such behavior was also observed experimentally in low temperature studies of
the peak longitudinal conductivity\cite{Wei1992}, see also Fig.~\ref{summary}.

\section{Diagrammatic approach to classical percolating transport}
\label{sec:diag}

\subsection{Systematic weak coupling expansion and extrapolation to the
percolation regime}

Our goal in this section is to discuss how classical percolation features in
quantum Hall transport can be captured analytically by a diagrammatic
approach\cite{FFC2011}, allowing to recover the percolation
Ansatz~(\ref{eq:scaling}) and accurate estimates of the critical exponent
$\kappa$ discussed in Sec.~\ref{hierachy}.
Building on earlier works\cite{Dreizer1972,Stroud1975} for the case of the local
conductivity tensor Eq.~(\ref{condTensor}), one can show by standard
techniques that the disorder averaged longitudinal macroscopic conductivity reads
\begin{equation}
\begin{pmatrix}
\sigma_{xx} & -\sigma_{xy}\\
\sigma_{xy} & \sigma_{xx}
\end{pmatrix}
=
\begin{pmatrix}
\sigma_{0} & -\big<\sigma_{H}\big>\\
\big<\sigma_{H}\big> & \sigma_{0}
\end{pmatrix}
+
\big<\hat\chi(\ve{r})\big>,
\end{equation}
where $\hat\chi(\ve{r})$ obeys the equation of motion:
\begin{equation}
\label{TAT:eq:chifinal}
 \hat\chi(\ve{r})=\delta\sigma(\ve{r})\hat\epsilon
+\delta\sigma(\ve{r})\int\!\!\! d^2 r'\,
\hat \epsilon\, \hat{\cal{G}}_0(\ve{r}-\ve{r'})\hat\chi(\ve{r'}) .
\end{equation}
We have introduced above the Hall conductivity fluctuations 
$\delta\sigma(\ve{r})\equiv\sigma_H(\ve{r})-\big<\sigma_H\big>$,
the antisymmetric $2\times2$ tensor $\hat\epsilon$, and the Green's 
function:
\begin{align}
[\hat{\cal{G}}_0]_{ij}(\ve{r})=\frac{\partial}{\partial r_i}
\frac{\partial}{\partial r_j}
\int\!\!\!\frac{d^2 p}{(2\pi)^2}\frac{e^{i\ve{p}\cdot\ve{r}}}
{\sigma_0|\ve{p}|^2+0^+}.
\label{eq:Green}
\end{align}

In previous analyses of Eq.~(\ref{TAT:eq:chifinal}), several methods were
proposed, such as a mean-field treatment\cite{Stroud1975}, lowest order
perturbation theory\cite{TRO2005} in powers of 
$\big<[\delta\sigma]^2\big>/\sigma_0^2$, or self-consistent Born 
approximation\cite{Dreizer1972}.
Clearly these approaches are insufficient to capture the critical percolation
behavior in the strong coupling limit $\sigma_0\to0$. However, the small
dissipation Ansatz~(\ref{eq:scaling}) ressembles the critical behavior typical
of phase transitions, and leads hope that Pad\'e extrapolation techniques
of a sufficiently high order perturbative calculation could bridge the gap
from weak ({\it i.e.} $\big<[\delta\sigma]^2\big> \ll \sigma_0^2$) to strong coupling
({\it i.e.} $\big<[\delta\sigma]^2\big>\gg \sigma_0^2$). The calculation
actually simplifies for the case of Gaussian fluctuations of the local Hall
conductivity $\delta\sigma(\ve{r})$, which applies to the highest temperature
regimes considered in Eq.~(\ref{sigmaH_classical}) and Eq.~(\ref{sigmaH_LL}).
By symmetry considerations, one finds that the Hall component is not affected 
in the high temperature regime, namely classical Hall's law $\sigma_{xy}= en/B$ holds.
By dimensional analysis, the longitudinal conductivity reads:
\begin{equation}
\sigma_{xx} 
= \sigma_0+
\sum_{n=1}^\infty a_{n}\frac{\langle\delta\sigma^2\rangle^n}{\sigma_0^{2n-1}}
\label{eq:series}
\end{equation}
with dimensionless coefficients $a_n$ collecting all diagrams of order $n$ in
perturbation theory in $\langle\delta\sigma^2\rangle/\sigma_0^2$.
The longitudinal conductivity $\sigma_{xx}$ thus receives non-trivial corrections that will 
lead to percolation effects in the limit $\sigma_0\to0$.

\begin{figure}
\includegraphics[width=0.7\linewidth]{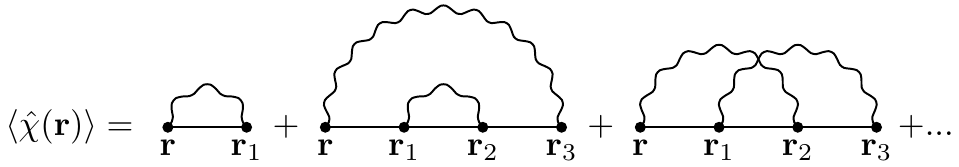}
\caption{Diagrammatic expansion in the case of Gaussian
fluctuations of the local conductivity. Wiggly lines are associated to disorder
averages, and solid lines to the Green's function Eq.~(\ref{eq:Green})}
\label{diagrams}
\end{figure}

The methodology to compute the large $\sigma_0$ expansion relies in iterating
Eq.~(\ref{TAT:eq:chifinal}) to the desired order, averaging over disorder
owing to the relation~(\ref{sigmaH_LL}), and evaluating the resulting multidimensional
integral, either analytically or numerically, see Fig.~\ref{diagrams}.
In order to simplify the calculations, we considered spatial correlations of
disorder of the form $ \langle\delta\sigma(\ve{r})\delta\sigma(\ve{r'})\rangle=
\langle\delta\sigma^2\rangle e^{-|\ve{r}-\ve{r'}|^2/\xi^2}$,
with correlation length $\xi$, allowing us to compute
the series Eq.~(\ref{eq:series}) up to sixth loop order\cite{FFC2011}, see 
Table~\ref{tab:values}.
\begin{table}[ht]
\tbl{Coefficients $a_n$ of the perturbative
series~(\ref{eq:series}) up to sixth loop order.}
{\begin{tabular}{ccc} \toprule
Order & Method & Coefficient $a_n$ \\
\colrule
1& Analytical & $\frac{1}{2}$ \\
2& Analytical & $\frac{1}{8}-\frac{1}{2}\log(2)$\\
3& Analytical & $0.2034560502$\\
4& Numerical & $-0.265\pm0.001$\\
5& Numerical & $0.405\pm0.001$\\
6& Numerical & $-0.694\pm0.001$\\
%\botrule
\end{tabular} \label{tab:values}}
\end{table}
Standard extrapolation techniques allow us to extract\cite{FFC2011} the 
estimate $\kappa=0.767\pm0.002$ for the critical exponent appearing in
Eq.~(\ref{eq:scaling}), quite close to the previously conjectured
value\cite{SH1994,PS1995,FS1995,Isichenko} $\kappa=10/13\simeq0.769$.

\subsection{High temperature crossover function for $\sigma_{xx}^\mathrm{peak}$} 

We finally provide a simple scaling function describing the crossover from
the high temperature regime above the cyclotron energy $T\gg \hbar\omega_c$
to the intermediate situation $v\ll T\ll \hbar\omega_c$, where Gaussian
fluctuation of the local Hall conductivity still arise, see Eq.~(\ref{sigmaH_LL}).
From this expression, we can connect the typical fluctuations of the Hall conductivity to
the width $v=\sqrt{\big<[V({\bf r})]^2\big>}$ of the disorder distribution:
$\sqrt{\big<[\delta\sigma({\bf r})]^2\big>} =
{\displaystyle \frac{e^2}{h}} v
\Bigg| {\displaystyle \sum_{m=0}^{\infty}} n_F'(E_m-\mu) \Bigg|
$,
so that the high temperature crossover function reads
from the scaling Ansatz~(\ref{eq:scaling}):
\begin{equation}
\sigma_{xx} = \sigma_0^{1-\kappa}
\left| \frac{e^2}{h} v
\sum_{m=0}^{\infty} n_F'(E_m-\mu)
\right|^\kappa.
\label{sigmapeak1}
\end{equation}
Note that for the Gaussian model studied here, a dimensionless prefactor
in Eq.~(\ref{sigmapeak1}) happens\cite{FFC2011} to be quite close to 1,
and has not been written.
At temperatures such that $v\ll T\ll \hbar\omega_c$ and at the $\nu\to\nu+1$ 
plateau transition, {\it i.e.} for $\mu=\hbar\omega_c(\nu+1/2)$, we thus find:
\begin{equation}
\sigma_{xx}^\mathrm{peak} = \sigma_0^{1-\kappa}
\left| \frac{e^2}{h}\frac{v}{4T}\right|^\kappa,
\end{equation}
recovering expression~(\ref{sigmaxx_LL}) for the longitudinal conductivity
in the limit $v\ll T\ll \hbar\omega_c$.

We can alternatively re-express the sum over Landau levels in Eq.~(\ref{sigmapeak1}) by 
using Poisson summation formula\cite{CM2001} in the limit $T<\mu$, giving:
\begin{equation}
\sigma_{xx}= \sigma_0^{1-\kappa}\left[
\frac{e^2}{h} \frac{v}{\hbar\omega_c} 
\Bigg|1+\sum_{l=1}^{+\infty}
(-1)^l \cos\left(\frac{2\pi l\mu}{\hbar\omega_c}\right) 
\frac{\frac{4\pi^2 l k_B T}{\hbar\omega_c}}
{\mathrm{sinh}\left(\frac{2\pi^2 l k_B T}{\hbar\omega_c}\right)}\Bigg|
\right]^\kappa
\label{sigmapeak2}
\end{equation}
vindicating expression~(\ref{sigmaxx_classical}) for the peak longitudinal conductivity 
in the $T\gg \hbar\omega_c$ limit.
Either Eq.~(\ref{sigmapeak1}) or Eq.~(\ref{sigmapeak2}) can be used to extract
the critical exponent $\kappa$ from experimental data in the range of
temperatures near the cyclotron energy.

%%%%%%%%%%%%%%%%%%%%%%%%%%%%%%%%%%%%%%%%%%%%%%%%%%%%%%%%%%%%%%%%%%%%%%%%%%%%%%%%%%%%%%%%%%%
\section{Perspectives}

As a conclusion, we list several issues that could be addressed in further
developments of the present work.
\begin{itemize}
\item What is the magnetic field behavior of $\sigma_{xx}$ at high temperature? 
\item Can one extract reliably the classical exponent $\kappa$ from experiments?
\item Are the exponents $\alpha$ and $\beta$ of the low temperature regime related to $\kappa$?
\item How do the finite probe currents affect the Hall plateau quantization?
\item What are the fundamental differences between 2DEGs and graphene\cite{CF2010}?
\item Is a more realistic description of electron-phonon conductivity\cite{Feng} needed?
\item Can one describe the crossover to Drude behavior at low magnetic
fields\cite{PEMW2001}?
\item Can one implement transport calculations using diagrammatic QMC \cite{Gull2011}?
\end{itemize}

%%%%%%%%%%%%%%%%%%%%%%%%%%%%%%%%%%%%%%%%%%%%%%%%%%%%%%%%%%%%%%%%%%%%%%%%%%%%%%%%%%%%%%%%%%%
\section*{Acknowledgments}
We thank S. Bera, A. Freyn, B. Piot, W. Poirier, M. E. Raikh, V. Renard and F. Schoepfer 
for stimulating discussions, and ANR ``Metrograph'' for financial support.


\begin{thebibliography}{80}

\bibitem{VonK1980}
K. Von Klitzing, G. Dorda, and M. Pepper, Phys. Rev. Lett. \textbf{45}, 494
(1980).

\bibitem{Prange}
R. E. Prange and S. M. Girvin, {\it The Quantum Hall Effect}, (Springer, New
York, 1987).

\bibitem{Hadju}
M. Janssen, O. Viehweger, U. Fastenrath, and J. Hadju, {\it Introduction to the
Theory of the Integer Quantum Hall Effect} (VCH, Germany, 1994).

\bibitem{Cage2005} J. Matthews and M.~E. Cage, J. Res. Natl. Inst. Stand.
Technol. \textbf{110}, 497 (2005).


\bibitem{Huckestein} B. Huckestein, Rev. Mod. Phys. \textbf{67}, 357 (1995).

\bibitem{Kettemann} B. Kramer, T. Ohtsuki and S. Kettemann, Phys. Rep.
\textbf{417}, 211 (2005).

\bibitem{EversRMP} F. Evers and A.~D. Mirlin, Rev. Mod. Phys. \textbf{80},
1355 (2008).


\bibitem{Wei1992} H.~P. Wei {\it et al.}, 
%S.~Y. Lin, D.~C. Tsui, and A.~M.~M. Pruisken,
Phys. Rev. B \textbf{45}, 3926 (1992).

\bibitem{Li2010} W. Li {\it et al.},
%J. S. Xia, C. Vicente, N. S. Sullivan, W. Pan, D. C. Tsui, L. N. Pfeiffer 
% and K. W. West, 
Phys. Rev. B \textbf{81}, 033305 (2010).

\bibitem{HSWIM2008} K. Hashimoto {\it et al.},
%, C. Sohrmann, J. Wiebe, T. Inaoka, F. Meier, Y. Hirayama, R. A. R\"omer, 
%R. Wiesendanger, and M. Morgenstern, 
Phys. Rev.  Lett. \textbf{101}, 256802 (2008).

\bibitem{SH1994} S.~H. Simon and B.~I. Halperin, Phys. Rev. Lett. \textbf{73},
3278 (1994).

\bibitem{PS1995} D.~G. Polyakov and B.~I. Shklovskii, Phys. Rev. Lett.
\textbf{74}, 150 (1995).

\bibitem{FS1995} M.~M. Fogler and B.~I. Shklovskii, Sol. State Comm.
\textbf{94}, 503 (1995).

\bibitem{FFC2011} M. Fl\"oser, S. Florens, and T. Champel,
Phys. Rev. Lett. {\bf 107}, 176806 (2011).

\bibitem{Isichenko} M.~B. Isichenko, Rev. Mod. Phys. \textbf{64}, 961 (1992).

\bibitem{GV1994} M.~R. Geller and G. Vignale, Phys. Rev. B \textbf{50}, 11714
(1994).

\bibitem{CFC2008} T. Champel, S. Florens and L. Canet, Phys. Rev. B
\textbf{38}, 125302 (2008).

\bibitem{CF2009} T. Champel and S. Florens, Phys. Rev. B {\bf 80}, 125322
(2009).

\bibitem{Feng} H. L. Zhao and S. Feng, Phys. Rev. Lett. \textbf{70}, 4134
(1993).

\bibitem{DR1994} A.~M. Dykhne and I.~M. Ruzin, Phys. Rev. B \textbf{50}, 2369
(1994).

\bibitem{Baranger} H. U. Baranger and A. D. Stone, Phys. Rev. B
\textbf{40}, 8169 (1989).

\bibitem{Dreizer1972} Y.~A. Dreizin and A.~M. Dykhne, Sov. Phys. JETP
\textbf{36}, 127 (1972).

\bibitem{Stroud1975} D. Stroud, Phys. Rev. B \textbf{12}, 3368 (1975).

\bibitem{TRO2005} C. Timm, M.~E. Raikh and F. von Oppen, Phys. Rev. Lett.
\textbf{94}, 036602 (2005).

\bibitem{CM2001} T. Champel and V.~P. Mineev, Philos Mag. B \textbf{81},
55 (2001).

\bibitem{CF2010} T. Champel and S. Florens, Physical Review B \textbf{82}, 045421 (2010).

\bibitem{PEMW2001} D.~G. Polyakov, F. Evers, A.~D. Mirlin and P. W\"olfle, 
Phys. Rev. B \textbf{64}, 205306 (2001).


\bibitem{Gull2011} E. Gull {\it et al.}, Rev. Mod. Phys. \textbf{83},
349 (2011).


\end{thebibliography}
\end{document}